\newcommand{\nn}{\nonumber}
\newcommand{\p}[1]{(\ref{#1})}
\documentstyle[11pt]{article}
\begin{document}

\thispagestyle{empty}
\begin{flushright}
JHU-TIPAC-96013\\
August 1996
\end{flushright}

\bigskip\bigskip\begin{center} {\bf
\Large{New Goldstone multiplet \\for partially broken
supersymmetry  }}
\end{center} \vskip 1.0truecm

\centerline{\bf Jonathan Bagger}
\vskip5mm
\centerline{and}
\vskip5mm
\centerline{\bf Alexander Galperin}
\vskip5mm
\centerline{Department of Physics and Astronomy}
\centerline{Johns Hopkins University}
\centerline{Baltimore, MD 21218, USA}

\vskip5mm

\bigskip \nopagebreak \begin{abstract}
\noindent
The partial spontaneous breaking of rigid $N=2$ supersymmetry
implies the existence of a massless $N=1$ Goldstone multiplet.
In this paper we show that the spin-$(1/2,1)$ Maxwell multiplet
can play this role.  We construct its full nonlinear transformation
law and find the invariant Goldstone action.  The spin-$1$ piece
of the action turns out to be of Born-Infeld type, and the full
superfield action is duality invariant.  This leads us to conclude
that the Goldstone multiplet can be associated with a $D$-brane
solution of superstring theory for $p=3$.  In addition, we find
that $N=1$ chirality is preserved in the presence of the
Goldstone-Maxwell multiplet.  This allows us to couple it to
$N=1$ chiral and gauge field multiplets.  We find that arbitrary
K\"ahler and superpotentials are consistent with partially
broken $N=2$ supersymmetry.

\end{abstract}

\newpage\setcounter{page}1
\begin{flushright}
{\it To the memory of Viktor I. Ogievetsky}
\end{flushright}

\section{Introduction.}

The spontaneous breaking of rigid supersymmetry gives rise to
a massless spin-1/2 Goldstone field, $\psi_\alpha(x)$ \cite{VA}.
When $N=2$ supersymmetry is broken to $N=1$, the Goldstone fermion
belongs to a massless multiplet of the unbroken $N=1$ supersymmetry.
One obvious Goldstone candidate is the $N=1$ chiral multiplet,
$(A+iB, \psi_\alpha, F+iG)$.  In ref.~\cite{BG} we used this
multiplet to construct a nonlinear realization of partially broken
$N=2$ supersymmetry.  We found that the complex spin-$0$ field
$A+iB$ is the Goldstone boson associated with the broken central
charge generator of $N=2$ supersymmetry; the complex auxiliary
field $F+iG$ parametrizes the coset $SU(2)/U(1)$ of the automorphism
group $SU(2)$.  The superthreebrane of Liu, Hughes and Polchinski
\cite{Pol} provides a different, but on-shell-equivalent
representation of the same Goldstone multiplet.

A second candidate Goldstone multiplet is the $N=1$ vector, or
Maxwell, multiplet, $(A_m, \psi_\alpha,$ $D)$.  In this case the
superpartners of the spin-1/2 Goldstone field are an abelian
gauge field $A_m$ and a real auxiliary field $D$.  We will show
that the Maxwell multiplet provides a second consistent Goldstone
multiplet for partially broken $N=2$ supersymmetry.  We will
construct its invariant action and its couplings to $N=1$ matter
fields.

Perhaps the most striking feature of the new Goldstone multiplet
is its unification of a Goldstone and a gauge field.  The theory
of Goldstone fields is based on the formalism of nonlinear
realizations, which is usually associated with finite-dimensional
groups \cite{CCWZ}, \cite{space}.  However, gauge fields can also
be interpreted as Goldstone fields associated with infinite-dimensional
symmetry groups \cite{IO}.  This suggests that the full symmetry of
the new multiplet is some infinite-dimensional extension of $N=2$
supersymmetry.  As we shall see, the gauge field $A_m$ has only
non-minimal interactions; in other words, the field appears only
via its field strength, so the gauge invariance is hidden.
Hence we can use the original formalism of \cite{CCWZ}, \cite{space}
to study the properties of the Goldstone-Maxwell multiplet.

This paper can be viewed as an outgrowth of an early attempt to
partially break $N=2$ supersymmetry \cite{BW}.  The problem of
ghost states is resolved by requiring the Goldstone multiplet
to be an irreducible representation of $N=1$ supersymmetry.
Recently, Antoniadis, Partouche and Taylor \cite{APT} constructed
a model with partially broken $N=2$ supersymmetry.  In their
model, the second supersymmetry is realized nonlinearly.  However,
their action involves an extra, massive, $N=1$ multiplet.  Our
approach is completely model-independent; if the extra
matter is integrated out, their action must reduce to ours.

Like the chiral Goldstone $N=1$ multiplet \cite{BG}, the
Goldstone-Maxwell multiplet has a superstring interpretation.
It is related to the recently discovered Dirichlet $p$-branes
\cite{D-branes}.  These objects are solutions of the superstring
equations of motion that can be viewed as dynamical membranes in
$(p+1)$-worldvolume space.  $D$-branes characteristically break
half of the superstring supersymmetries and involve a $(p+1)$-dimensional
gauge field with the Born-Infeld action.  Until now, only the
bosonic parts of the $D$-brane actions have been constructed.  We
propose that the Goldstone-Maxwell action, after eliminating the
auxiliary fields, is precisely the supersymmetric, gauge-fixed,
$D$-brane action for $p=3$ (in a flat background):  The gauge
field $A_m$ has a Born-Infeld action, and the full Goldstone-Maxwell
action is duality invariant.

This paper is organized as follows. In sect. 2 we review the
formalism of nonlinear realizations.  As we shall see, this
technique has an ambiguity when applied to $N=2$ supersymmetry:
dimensionless invariants can be used to modify the covariant
derivatives and the covariant constraints.  However, requiring
consistency of the constraints fixes the ambiguity.  In sect. 3
we find a set of consistent constraints, to third order in
the Goldstone fields.  We then solve the constraints in terms
of the ordinary $N=1$ Maxwell multiplet, and derive the broken
supersymmetry transformations of the Goldstone-Maxwell multiplet
to second order in fields.  In sect. 4 we present the full
nonlinear transformation law and derive the invariant action
for the Goldstone-Maxwell multiplet.  Surprisingly, we find
that the gauge field is governed by the Born-Infeld action,
and that the full action is invariant under a superfield duality
transformation.  In sect 5. we show that $N=1$ chirality is
preserved in the presence of the Goldstone-Maxwell multiplet.
This allows us to generalize the K\"ahler potential to the case
of partially broken $N=2$ supersymmetry.  It also permits us to
construct $N=2$ extensions of the general superpotential for chiral
$N=1$ superfields, as well as the kinetic and Fayet-Iliopoulos
terms for $N=1$ gauge superfields.  Section 6 contains concluding
remarks.

\section{Nonlinear realizations.}

In this section we review basics of nonlinear realizations as applied
to $N=2$ supersymmetry.  We begin with the $N=2$ supersymmetry algebra,
\begin{eqnarray}
\label{algebra}
\{Q_\alpha, \bar Q_{\dot\alpha}\}\ =
\ 2\sigma^a_{\alpha\dot\alpha}P_a\ , &&
\{S_\alpha, \bar S_{\dot\alpha}\}\ =
\ 2\sigma^a_{\alpha\dot\alpha}P_a\ , \\
\{Q_\alpha, S_\beta\}\  =\ 0\ , &&
\{Q_{\alpha}, \bar S_{\dot\alpha}\}\ =\ 0\ ,
\nonumber
\end{eqnarray}
where $Q_\alpha$ and $S_\alpha$ are the two supersymmetry generators
and $P_a$ is the four-dimensional momentum operator.  In what follows,
we take $Q_\alpha$ to be the unbroken $N=1$ supersymmetry generator,
and $S_\alpha$ to be its broken counterpart.

Following the formalism of nonlinear realizations \cite{CCWZ},
\cite{space}, we consider the coset space $G/H$, where $G$ is the
$N=2$ supersymmetry group and $H=SO(3,1)$ is its Lorentz subgroup.
We parametrize the coset element $\Omega$ as follows,
\begin{equation}
\label{real_parametrization}
\Omega\ =\ \exp i(x^aP_a +\theta^\alpha Q_\alpha +\bar\theta_{\dot\alpha}
\bar Q^{\dot\alpha})
\exp i(\psi^\alpha S_\alpha +\bar\psi_{\dot\alpha}\bar S^{\dot\alpha})\ .
\end{equation}
Here $x,\ \theta$ and $\bar\theta$ are coordinates of $N=1$
superspace, while $\psi^\alpha=\psi^\alpha(x, \theta, \bar\theta)$
and its conjugate $\bar\psi_{\dot\alpha}=\bar\psi_{\dot\alpha}(x,
\theta, \bar\theta)$ are Goldstone $N=1$ superfields of dimension
$-1/2$.  Note that these superfields are reducible; they contain
spins up to 3/2.  In the next section we will reduce the
representations by imposing $N=2$ covariant irreducibility
constraints.

The group $G$ acts on the coset space by left multiplication
\begin{equation}
\label{left_multiplication}
g\Omega\ =\ \Omega' h \;\;\;\;\;\;(g \in G, \; h\in H)\ .
\end{equation}
In particular, under $S$-supersymmetry, with $g=\exp i(\eta S+
\bar\eta\bar S)$, this implies
\begin{eqnarray}
x'^a\ &=&\ x^a\ +\ i(\eta\sigma^a\bar\psi
-\psi\sigma^a\bar\eta) \ ,\nn\\
\; \theta'\ &=&\ \theta \ ,\nn\\
\bar\theta'\ &=&\  \bar\theta\ ,
\end{eqnarray}
and
\begin{eqnarray}
\psi'^\alpha(x', \theta', \bar\theta')\ &=&\ \psi^\alpha(x,
\theta, \bar\theta)\ +\ \eta^\alpha\ ,\nn\\
\bar\psi'_{\dot\alpha}(x', \theta', \bar\theta')\ &=&
\ \bar\psi_{\dot\alpha}(x,
\theta, \bar\theta) \ +\ \bar\eta_{\dot\alpha}\ .
\end{eqnarray}
The Cartan 1-form $\Omega^{-1}d\Omega$,
\begin{equation}
\label{1-form}
\Omega^{-1}d\Omega\ =\ i \left[\omega^a(P)P_a+\omega^\alpha(Q)Q_\alpha
+ \bar\omega_{\dot\alpha}(\bar Q)\bar Q^{\dot\alpha}
+\omega^\alpha(S)S_\alpha +
\bar\omega_{\dot\alpha}(\bar S)\bar S^{\dot\alpha}\right]\ ,
\end{equation}
defines covariant $N=1$ superspace coordinate differentials
\begin{eqnarray}
\omega^a(P)\ &=&\ dx^a\ +\ i(d\theta\sigma^a\bar\theta +
d\bar\theta\bar\sigma^a\theta
+d\psi\sigma^a\bar\psi + d\bar\psi\bar\sigma^a\psi)\ , \nn\\
\omega^\alpha(Q)\ &=&\ d\theta^\alpha\ , \nn\\
\bar\omega_{\dot\alpha}(\bar Q)\ &=&
\ d\bar\theta_{\dot\alpha}\ ,
\end{eqnarray}
and covariant Goldstone one-forms
\begin{eqnarray}
\omega^\alpha(S)\ &=&
\ d\psi^\alpha\ , \nn\\
\bar\omega_{\dot\alpha}(\bar S)\ &=& \ d\bar\psi_{\dot\alpha}\ .
\end{eqnarray}

The supervielbein matrix $E_M{}^A$ is found by expanding the
one-forms $\omega^A \equiv (\omega^a(P),$ $\omega^\alpha(Q),$
$\bar\omega_{\dot\alpha}(\bar Q))$ with respect to the $N=1$
superspace coordinate differentials $dX^M = (dx^m, d\theta^\mu,
d\bar\theta_{\dot\mu})$,
\begin{equation}
\omega^A\ =\ dX^M E_M{}^A\ .
\label{vielbein}
\end{equation}
In a similar fashion, the covariant derivatives of the Goldstone
superfield $\psi^\alpha$ are found by expanding $\omega^\alpha(S)
= \omega^A {\cal D}_A\psi^\alpha$, which implies ${\cal D}_A\psi^\alpha =
E^{-1}_A{}^M\partial_M\psi^\alpha$.  These covariant derivatives
can be explicitly written as follows,
\begin{eqnarray}
\label{derivatives}
{\cal D}_a\ &=&\ \omega^{-1}_a{}^m\partial_m\ , \nn \\
{\cal D}_\alpha\ &=&\ D_\alpha
\ -\ i(D_\alpha\psi\sigma^a\bar\psi +
 D_\alpha\bar\psi\bar\sigma^a\psi)\omega^{-1}_a{}^m\partial_m\ ,  \\
\bar{\cal D}_{\dot\alpha}\ &=&\ \bar D_{\dot\alpha}\ -
 \ i(\bar D_{\dot\alpha}\psi\sigma^a\bar\psi\ +
\bar D_{\dot\alpha}\bar\psi\bar\sigma^a\psi)\omega^{-1}_a{}^m
\partial_m \ ,\nn
\end{eqnarray}
where $\omega_m{}^a \equiv  \delta_m^a + i(\partial_m\psi\sigma^a\bar\psi +
\partial_m\bar\psi\bar\sigma^a\psi)$ and $D_\alpha, \bar D_{\dot\alpha}$
are the ordinary flat $N=1$ superspace spinor derivatives.

$N=1$ matter superfields $\Phi(x, \theta, \bar\theta)$ transform
as follows under the full group $G$,
\begin{equation}
\Phi'(x', \theta', \bar\theta')\ =\ R(h)\Phi(x, \theta,\bar\theta)\ ,
\end{equation}
where $R(h)$ is a matrix in a representation of the
stability subgroup, $H=SO(3,1)$.  Since there is no $H$-connection
in the right-hand side of \p{1-form}, the covariant derivatives of
the matter and Goldstone $N=1$ superfields are identical.

In what follows we will need the algebra of the covariant derivatives.
This algebra can be worked out with the help of \p{derivatives}:
\begin{eqnarray}
\label{cov_algebra}
\{ {\cal D}_\alpha,  {\cal D}_\beta \} \ & = &
\ -\ 2i({\cal D}_\alpha\psi^\gamma
{\cal D}_\beta\bar\psi^{\dot\gamma} +
{\cal D}_\beta\psi^\gamma
{\cal D}_\alpha\bar\psi^{\dot\gamma})
\sigma^a_{\gamma\dot\gamma}{\cal D}_a\ , \nn\\
\{ {\cal D}_\alpha,  \bar{\cal D}_{\dot\beta} \} \ & = &
\ 2i\sigma^a_{\alpha\dot\beta}{\cal D}_a -2i({\cal D}_\alpha\psi^\gamma
\bar{\cal D}_{\dot\beta}\bar\psi^{\dot\gamma} + \bar{\cal
D}_{\dot\beta}\psi^\gamma
{\cal D}_{\alpha}\bar\psi^{\dot\gamma} ) \sigma^a_{\gamma\dot\gamma}{\cal D}_a\
,  \\
\left[ {\cal D}_\alpha,  {\cal D}_a \right] \ & = &
\ -\ 2i ({\cal D}_\alpha \psi^\gamma
{\cal D}_a \bar\psi^{\dot\gamma} +
{\cal D}_a\psi^\gamma
{\cal D}_\alpha \bar\psi^{\dot\gamma})
\sigma^b_{\gamma\dot\gamma}{\cal D}_b\ . \nn
\end{eqnarray}

An important feature of this formalism is the existence of two
dimensionless invariants, $\bar{\cal D}_{\dot\alpha}\psi_\alpha$
and ${\cal D}_\alpha\psi_\beta$ (together with their complex
conjugates).  These invariants render the standard formalism of
nonlinear realizations somewhat ambiguous.  For example, one can
multiply the covariant derivative ${\cal D}_\alpha$ by one function
of these invariants, or shift ${\cal D}_\alpha\psi_\beta$ by another.
This ambiguity will prove important in the next section; the
reason for it and the way to overcome it will be discussed in
the conclusions.

\section{Consistent covariant constraints.}

The Goldstone superfield $\psi^\alpha(x, \theta, \bar\theta)$ is
a reducible representation of unbroken $N=1$ supersymmetry with
highest spin $3/2$.  It contains the spin-$(1, 1/2)$ Maxwell
multiplet, but it also contains ghosts.  The only way to
eliminate the ghosts is to impose appropriate irreducibility
constraints on the Goldstone superfield.

In this section we will find a set of consistent, $N=2$ covariant
constraints which reduce $\psi^\alpha$ to the $N=1$ Maxwell
multiplet.  The elucidation of the proper constraints is complicated
by the dimensionless invariants discussed in the previous section.
Therefore we will adopt a perturbative approach and present a set
of constraints that are consistent to the third order in the
Goldstone fields.

As is well-known (see, e.g. \cite{BWbook}), the Maxwell multiplet
is described by a chiral $N=1$ field strength $W_\alpha$ of
dimension $3/2$
\begin{equation}
\bar D_{\dot\alpha}W_\alpha\ = 0\ .
\label{linearized_a}
\end{equation}
The superfield $W_\alpha$ satisfies the irreducibility
constraint
\begin{equation}
D^\alpha W_\alpha + \bar D_{\dot\alpha} \bar W^{\dot\alpha}
\ =\ 0\ .
\label{linearized_b}
\end{equation}
The second constraint \p{linearized_b} must also satisfy a
consistency condition:  its left-hand side must vanish under
$D^2$ (and $\bar D^2$).

The two constraints are solved by $W_\alpha = i \bar D^2D_\alpha
V$, where $V(x, \theta, \bar\theta)$ is the Maxwell prepotential.
The field $V$ is defined modulo chiral gauge transformations,
$\delta V = i(\Lambda - \bar\Lambda)$, with $\bar D_{\dot\alpha}
\Lambda = 0$.

To lowest order, we can identify
\begin{equation}
\psi_\alpha \big\vert_{\rm lin}\ =\ \kappa^2 W_\alpha\ ,
\end{equation}
where the constant $\kappa$ (of dimension $-2$) is the scale of
$S$-supersymmetry breaking.  In what follows, we set $\kappa = 1$.

Our aim is to generalize \p{linearized_a}, \p{linearized_b} to
obtain a set of constraints that are covariant under $N=2$
supersymmetry.  The new constraints must be consistent and
reduce to \p{linearized_a},\p{linearized_b} in the linearized
approximation.

We begin by generalizing\footnote{Equation
\p{chirality_constraint} was first discussed in similar
context in \cite{BW}.}  \p{linearized_a}
\begin{equation}
\bar{\cal D}_{\dot\beta}\psi_\alpha\ =\ 0\ .
\label{chirality_constraint}
\end{equation}
Note that the right-hand side of this equation can, in principle,
involve any power of the dimensionless invariants ${\cal D}^\alpha
\psi_\beta$ and $\bar{\cal D}_{\dot\beta}\psi_\alpha$.  However,
it is easy to see that \p{chirality_constraint} is consistent
as it stands. Indeed, Lorentz covariance implies that any terms
on the right side \p{chirality_constraint} must be at least linear
in $\bar{\cal D}\psi$ and ${\cal D}\bar\psi$.  Hence the most
general modification of
\p{chirality_constraint} has the form $\bar{\cal D}_{\dot\alpha}
\psi_\alpha = M_{\alpha\dot\alpha}{}^{\beta\dot\beta}
\bar{\cal D}_{\dot\beta}\psi_\beta + N_{\alpha\dot\alpha}{}
^{\beta\dot\beta} {\cal D}_\beta\bar\psi_{\dot\beta}$, where
the matrices $M$ and $N$ are at least linear in the Goldstone
fields.  This equation, together with its conjugate, imply
\p{chirality_constraint}.  An important corollary of this
result is the fact that $N=1$ chirality is preserved in the
Goldstone background,
\begin{equation}
\label{chirality_preservation}
\{ {\cal D}_\alpha, {\cal D}_\beta \}\ =
\ \{ \bar{\cal D}_{\dot\alpha}, \bar{\cal D}_{\dot\beta} \}
\ =\ 0\ .
\end{equation}
We will discuss the geometrical meaning of covariant chirality
in sect. 5.

We now turn to \p{linearized_b}. The simplest, most naive
generalization, ${\cal D}^\alpha\psi_\alpha + \bar{\cal D}_{\dot
\alpha}\bar\psi^{\dot\alpha} = 0$, is not consistent at order
${\cal O}(\psi^3)$.  Applying $\bar{\cal D}^2$ to the left-hand
side gives
\begin{equation}
\bar{\cal D}^2 ( {\cal D}^\alpha\psi_\alpha + \bar{\cal
D}_{\dot\alpha}\bar\psi^{\dot\alpha} )\ =
\ 4 D\psi\partial^{\dot\alpha\alpha}\psi^\beta
\partial_{\beta\dot\alpha}\psi_\alpha\ +
\ 8 D^\alpha\psi^\beta \partial^{\dot\alpha\gamma}\psi_\gamma
\partial_{\beta\dot\alpha}\psi_\alpha\ +\ {\cal O}(\psi^5)\ .
\end{equation}
It is remarkable that there exists the following $N=2$ covariant
generalization of the Maxwell constraints,
\begin{equation}
\label{covariant_b}
{\cal D}^\alpha\psi_\alpha\ +\ \bar{\cal
D}_{\dot\alpha}\bar\psi^{\dot\alpha}\ -
\ {1\over 2}{\cal D}^\alpha\psi^\beta{\cal D}_\beta\psi_\alpha
{\cal D}^\gamma\psi_\gamma\ -\ {1\over 2}
\bar{\cal D}^{\dot\alpha}\bar\psi^{\dot\beta}
\bar{\cal D}_{\dot\beta}\bar\psi_{\dot\alpha}
\bar{\cal D}_{\dot\gamma}\bar\psi^{\dot\gamma}
\ = \ {\cal O}(\psi^5)\ .
\end{equation}
This constraint is consistent to order ${\cal O}(\psi^3)$, in the sense
that $\bar{\cal D}^2 ({\rm l.h.s. \p{covariant_b} }) = {\cal O}(\psi^5).$

The ambiguity of the standard nonlinear realization
is completely fixed by the consistency requirements.  In fact,
higher-order terms can be added to the left-hand side of \p{covariant_b}
to make it consistent to all orders.  The structure of these
higher-order corrections is likely to be related to hidden
symmetries of the Goldstone-Maxwell multiplet.

The consistent covariant constraints \p{chirality_constraint}
and \p{covariant_b} can be solved in terms of the $N=1$ Maxwell
field strength $W_\alpha$,
\begin{equation}
\label{solution}
\psi_\alpha\ =\ W_\alpha \ +\ {1\over 4} \bar D^2(\bar W^2) W_\alpha
\ -\ i W^\beta\bar W^{\dot\beta}\partial_{\beta\dot\beta}W_\alpha
\ +\ {\cal O}(W^5)\ ,
\end{equation}
where $W^2 = W^\alpha W_\alpha$ and $\bar W^2 = \bar W_{\dot\alpha}
\bar W^{\dot\alpha}$.

In what follows an important role is played by the nonlinear
transformations of $W_\alpha$ under the second supersymmetry.
To find them, let us first consider the form-variation of
$\psi_\alpha$ under $S$-supersymmetry,
\begin{eqnarray}
\delta^*\psi_\alpha\ &\equiv&\ \psi'_\alpha(x, \theta, \bar\theta)
\ -\ \psi_\alpha(x, \theta, \bar\theta)\nn\\
&=&\ \eta_\alpha\ -\ i
(\eta^\beta\bar\psi^{\dot\beta} \ -\ \psi^\beta\bar\eta^{\dot\beta})
\partial_{\beta\dot\beta}\psi_\alpha\ .
\end{eqnarray}
For the Maxwell field strength $W_\alpha$, this implies
\begin{equation}
\label{second_susy_on_W}
\delta^*W_\alpha\ =\ \eta_\alpha - {1\over 4} \bar D^2(\bar W^2)
\eta_\alpha\ -\ i\partial_{\alpha\dot\alpha}(W^2)
\bar\eta^{\dot\alpha}\ +\ {\cal O}(W^4)\ .
\end{equation}
Note that this transformation preserves the defining linear
constraints \p{linearized_a}, \p{linearized_b}.  The
corresponding Maxwell prepotential transformation is
\begin{equation}
\label{second_susy_on_V}
\delta^*V\ =\ {i\over 4}(\bar\theta^2 +\bar W^2)\theta\eta
\ -\ {i\over 4}(\theta^2 + W^2)\bar\theta\bar\eta\ +\ {\cal O}(W^4)\ .
\end{equation}
The commutator of two such transformations reduces to an
ordinary translation (plus a gauge transformation), as required
by the algebra \p{algebra}.

Using \p{second_susy_on_W}, we can find the $N=2$ invariant
Goldstone-Maxwell action (to order $W^6$),
\begin{equation}
\label{goldstone_action}
S_{\rm goldst}\ =\ {1\over 4}
\int d^4x d^2\theta W^2\ +\ {1\over 4}\int d^4x d^2\bar\theta
\bar W^2\ +\ {1\over 8}  \int d^4x d^4\theta W^2\bar W^2  \ +
\ {\cal O}(W^6).
\end{equation}
The gauge field contribution to this action has the form
\begin{equation}
\label{gauge_action}
(S_{\rm goldst})\bigg\vert_{\rm gauge}\ =\ \int
d^4 x \left[ -{1\over 4}F_{mn}F^{mn}
-{1\over 32}(F_{mn}F^{mn})^2 \ +\ {1\over 8}F_{mn}F^{nk}F_{kl}F^{lm}\right]
\ +\ {\cal O}(F^6)\ .
\end{equation}
The action \p{gauge_action} coincides with the
expansion of the Born-Infeld action
\begin{equation}
\label{BI_action}
S_{\rm BI}\ = \ -\ \int d^4 x\ \sqrt{ - \det(\eta_{mn} + F_{mn})}
\ .
\end{equation}
In the next section we will see that this is not an accident;
the full nonlinear action for the gauge field is precisely
that of Born and Infeld \cite{Born Infeld}.

\section{The Goldstone-Maxwell multiplet.}

\subsection{The nonlinear transformation law.}

In this section we will extend the results of the previous
section to all orders.  Instead of generalizing the constraints
\p{chirality_constraint}, \p{covariant_b}, we will work directly
with the $N=1$ Maxwell superfield $W_\alpha$.  We stress that
all results of this  section are nonperturbative.

We begin with the full nonlinear transformation law for $W_\alpha$.
To preserve the defining constraints \p{linearized_a}, \p{linearized_b}
it must have the form
\begin{equation}
\label{ansatz}
\delta^*W_\alpha\ =\ \eta_\alpha\ -\ {1\over 4}\bar D^2\bar X\eta_\alpha
\ -\ i\partial_{\alpha\dot\alpha}X\bar\eta^{\dot\alpha}\ ,
\end{equation}
where $X$ is a chiral $N=1$ superfield which satisfies $\bar
D_{\dot\alpha} X = 0$. The commutator of two such transformations
obeys the $N=2$  algebra \p{algebra} if $X$ transforms as
\begin{equation}
\label{deltaX}
\delta^*X\ =\ 2W^\alpha\eta_\alpha\ .
\end{equation}
Note that the commutator of two such transformations gives the
correct algebra.

The following recursive expression for $X$ is chiral and has the
required transformation properties:
\begin{equation}
\label{recursive}
X\ =\ {W^2 \over 1 - {1\over 4}\bar D^2\bar X}\ .
\end{equation}
We will not derive this equation since it was guessed.  However,
once found, it can be justified by its consistency with \p{ansatz}
and  \p{deltaX}.\footnote{Note that there can exist only one superfield
$X$ with the required properties:  given two such superfields, $X$
and $X'$, $X-X'$ is invariant under $S$-supersymmetry.  No such
invariant of dimension 1 can be built from $W_\alpha$, except for
a constant.  The constant part of $X$ is fixed by requiring $X$
vanish at $W_\alpha = 0$.}

Equation \p{recursive} can be used to expand $X$ in powers of $W^2$
and its derivatives,
\begin{equation}
\label{perturbation}
X\ =\ W^2\ +\ {1\over 4}W^2 \bar D^2(\bar W^2)\ +
\ {1\over 16}W^2\left[(\bar D^2\bar W^2)^2\ +\ \bar D^2( \bar W^2\bar D^2\bar
W^2)  \right] \ +\ \ldots
\end{equation}
More importantly, it can also be used to find an explicit expression
for $X$.  To this end, we transform \p{recursive} in the following
way,
\begin{eqnarray}
\label{transformed}
X\ &=&\ W^2\ +\ {W^2 \over 4} {\bar D^2\bar X \over 1 -
{1\over 4}\bar D^2\bar X}\nonumber\\
&=&\ W^2\ +\ {1 \over 4}\bar D^2\left[ {W^2\bar W^2 \over
(1 - {1\over 4} D^2 X )(1 - {1\over 4}\bar D^2\bar X)}\right]\ .
\end{eqnarray}
We note that the numerator in the square brackets involves the
squares of the anticommuting spinor superfields $W_\alpha$ and
$\bar W_{\dot\alpha}$.  Since $W_\alpha W_\beta W_\gamma = 0$
and $\bar W_{\dot\alpha}\bar W_{\dot\beta} \bar W_{\dot\gamma}
= 0$, the terms in the denominator which contain an
undifferentiated $W$ or $\bar W$ must vanish.  This implies
that $D^2X$ enters the denominator only in the following
``effective" form,
\begin{equation}
\label{D^2Xeffective}
(D^2X)_{\rm eff}\ =\ { D^2W^2 \over 1 -{1\over 4} (\bar D^2\bar
X)_{\rm eff}}\ .
\end{equation}
This equation, together with its complex conjugate, gives rise
to a quadratic equation for $(D^2X)_{\rm eff}$, with the
solution\footnote{The second solution does not vanish at $W=0$
and should be discarded.}
\begin{equation}
\label{D^2Xsolution}
(D^2X)_{\rm eff}\ =\ 2\ +\ B\ -\ 2 \sqrt{1-A+{1\over 4}B^2}\ ,
\end{equation}
where
\begin{eqnarray}
\label{ABdefinition}
A\ &=&\ {1\over 2}(D^2W^2 + \bar D^2\bar W^2)\ , \nn\\
B\ &=&\ {1\over 2}(D^2W^2 - \bar D^2\bar W^2)\ .
\end{eqnarray}
Substituting this into \p{transformed}, we find an explicit
expression for $X$,
\begin{equation}
\label{Xexplicit}
X\ =\ W^2\ +\ {1\over 2}\bar D^2\left[ {W^2\bar W^2 \over 1 - {1\over 2}
A\ +\ \sqrt{1-A+{1\over 4}B^2}}\right]\ .
\end{equation}

\subsection{The action.}

The superfield $X$ plays a second important role:  it is also a
chiral density for the invariant Goldstone-Maxwell action. Indeed,
the transformation property \p{deltaX} implies that the chiral
integral
\begin{equation}
\label{Xintegral}
\int d^4xd^2\theta\ X
\end{equation}
is invariant under $N=2$ supersymmetry. The $Q$-supersymmetry is
manifest in \p{Xintegral}, while the $S$-invariance follows from
the fact that $\int d^4xd^2\theta W^\alpha\eta_\alpha$ is a surface
term.  The Goldstone-Maxwell action is nothing but the real part
of the invariant  \p{Xintegral},\footnote{The imaginary part of \p{Xintegral}
reduces to a surface term.}
\begin{eqnarray}
\label{Xaction}
S_{\rm GM}\ &=&\ {1\over 4} \int d^4xd^2\theta\ X\ +\ {1\over 4} \int
d^4xd^2\bar\theta\ \bar X \nn \\
&=&\ {1\over 4} \int d^4xd^2\theta\ W^2 \ +\ h.c.\ +\ {1\over 4}
\int d^4xd^2\theta d^2\bar\theta
\ {W^2\bar W^2 \over 1 - {1\over 2} A +
\sqrt{1-A+{1\over 4}B^2}}\ .
\label{GMaction}
\end{eqnarray}
By construction, this action is invariant under $N=2$ supersymmetry,
where the second supersymmetry is given by \p{ansatz}, \p{Xexplicit},
and \p{ABdefinition}.  It is written in terms of the $N=1$ Maxwell
multiplet field strength $W_\alpha$ and its derivatives.

Physically, the action \p{GMaction} describes nonminimal couplings
of massless spin-$1/2$ and spin-$1$ particles, with first and
second order equations of motion, respectively.  It does not
involve any ghost states.

It is instructive to analyse the
bosonic part of the action.  To this end we set the fermionic
field $W_\alpha|_{\theta =0} = 0$, and use the identities
\begin{eqnarray}
\label{DWidentity}
D^\alpha W_\beta &=& {1\over 4}(\sigma^{mn}
F_{mn})_\beta{}^\alpha\ +\ {i\over 4} \delta_\beta^\alpha D \nn\\
D^2W^2 &=& -{1\over 2} F_{mn} F^{mn}\ -\ {i\over 2} F_{mn} \tilde F^{mn}
\ + \ D^2 \ ,
\end{eqnarray}
which hold at $W_\alpha|_{\theta=0} = 0$.
Since Grassmann integration is equivalent to differentiation,
\p{GMaction} and \p{DWidentity} imply that the real field $D$ enters
the bosonic action in a bilinear way.  Therefore on shell, $D=0$, and
the gauge field strength $F_{mn}$ contains all the bosonic degrees
of freedom.

The action for the gauge field $F_{mn}$ can be written as
\begin{equation}
S_{{\rm bosonic}}\ =\ \int d^4x \left[1-
\left(1 + {1\over 2}F_{mn}F^{mn} + {1\over 8}(F_{mn}F^{mn})^2 -
{1\over 4}F_{mn}F^{nk}F_{kl}F^{lm}\right)^{1/2} \right]\ .
\end{equation}
Since in four dimensions
\begin{equation}
- \det(\eta_{mn} + F_{mn})\ =\ 1\ +\ {1\over 2}F_{mn}F^{mn}
\ +\ {1\over 8}(F_{mn}F^{mn})^2\ -\ {1\over 4}F_{mn}F^{nk}
F_{kl}F^{lm}\ ,
\end{equation}
this action coincides (up to additive constant) with the Born-Infeld
action \p{BI_action}.  It is remarkable that the
Born-Infeld form of the gauge field action is dictated by the partially
broken $N=2$ supersymmetry.  Since the gauge field is a superpartner of
the Goldstone fermion, this hints strongly that a Goldstone-type
symmetry underlies the Born-Infeld action.

We should mention that the action \p{GMaction} was first constructed
in \cite{CF} as an $N=1$ generalization of the Born-Infeld
action.\footnote{We are grateful E. Ivanov for introducing us
to this paper.}  As pointed out in \cite{CF}, the $N=1$ supersymmetry
is not sufficient to fix at the action \p{GMaction}.  Indeed, one
can modify  the $d^4\theta$ part of the $N=1$ Lagrangian by
replacing $D^2W^2 \rightarrow D^2W^2 + a (D^\alpha W_\alpha)^2$,
where $a$ is any number.  This clearly does not change the
Born-Infeld  form of the gauge field action.  Note, however, that
this modification is not consistent with the transformations
\p{ansatz}, \p{Xexplicit}.  It is the second, nonlinear supersymmetry
which unambiguously defines the form of the Goldstone-Maxwell action.

\subsection{Duality.}

We now turn to the duality properties of the Goldstone-Maxwell action.
Let us first recall that the Born-Infeld action possesses a certain
duality invariance \cite{T}.  The duality stems from the fact that the
action involves the field strength only; therefore one can relax the
Bianchi identity $\epsilon^{mnkl} \partial_n F_{kl}=0$ by including a
Lagrange multiplier term
\begin{equation}
\label{relaxedLM}
S_{\rm BI}(F_{mn})\ \rightarrow
\ S_{\rm BI}(F_{mn})\ +\ {1\over 2}\int d^4x\ \tilde A_m
\epsilon^{mnkl}\partial_n F_{kl}\ ,
\end{equation}
where $\tilde A_m$ is the multiplier field.
If one varies this action with respect to $F_{mn}$, and substitutes
back the result, one recovers the Born-Infeld action for
$\tilde A_m$ itself.

The Goldstone-Matter action \p{GMaction} enjoys a similar
self-duality.  This can be seen
as follows.  We first relax the Bianchi identity
\p{linearized_b} by adding a superfield Lagrange multiplier
term to \p{GMaction}
\begin{equation}
\label{relaxedGM}
S_{\rm GM}(W)\ \rightarrow
\ S_{\rm GM}(W)\ +\ {i\over 2}\int d^4x d^2\theta
\ \widetilde W^\alpha W_\alpha\ -\ {i\over 2}\int d^4x d^2\bar\theta
\ \widetilde{\overline{W}}_{\dot\alpha} \overline W^{\dot\alpha}\ .
\end{equation}
Here $\widetilde W^\alpha$ is an $N=1$ Maxwell multiplet which
serves as an $N=1$ Lagrange multiplier, and $W_\alpha$ is an
arbitrary chiral $N=1$ superfield.  Varying with respect to
$\widetilde W^\alpha$ reimposes the Maxwell constraint on
$W_\alpha$, while varying with respect to $W^\alpha$ gives rise
to the Goldstone-Maxwell action for the field $\widetilde W^\alpha$.

To see how this works, let us first vary \p{relaxedGM} with respect
to $W_\alpha$.  This gives
\begin{equation}
\label{W_thru_W}
W^\alpha\ =\ -\ i \widetilde
W^\alpha (1 - {1\over 4} \bar D^2\bar X)\, t^{-1}\ ,
\end{equation}
where $t$ satisfies to the recursive equation
\begin{equation}
\label{t_equation}
t\ =\ 1\ -\ {1\over 4}\bar D^2 (\widetilde{\overline{W}}^2
\bar t^{-1})\ .
\end{equation}
We then substitute \p{W_thru_W} back into \p{relaxedGM}.  Let us
focus on the part of the action which is an integral over the full
$N=1$ superspace.  The trick of sect.~4 can be used to write $t$ and
$\bar D^2\bar X$ in the ``effective" form
\begin{eqnarray}
t_{{\rm eff}} &=& 1 \ -\ {\bar D^2\widetilde{\overline W}^2 \over
4\bar t_{{\rm eff}}}\ , \label{t_eff} \\[3mm]
(1 - {1\over 4} \bar D^2\bar X)_{{\rm eff}} &=& {t(\bar t - t +1)
\over \bar t + t -1}\Bigg|_{{\rm eff}}\ .
\nn
\end{eqnarray}
Solving \p{t_eff} for $t$ and substituting back into the action, we
recover the Goldstone-Maxwell action \p{GMaction} for the $N=1$
Maxwell superfield $\widetilde W_\alpha$.

In this section we established that the Goldstone-Maxwell action possesses
partially broken $N=2$ supersymmetry, that it is self-dual, and that its
bosonic part reduces to the Born-Infeld action.  These are exactly the
properties that are expected from a supersymmetric $D$-brane action
\cite{D-branes}.  Thus we may conclude that \p{GMaction} is, in fact,
the gauge-fixed $D$-brane action in a flat background (after the
auxiliary fields are eliminated).

\section{$N=1$ matter and the Goldstone-Maxwell multiplet.}

\subsection{Chirality.}

The chirality constraint \p{chirality_constraint} and integrability
of the covariant spinor derivatives \p{chirality_preservation} allow
us to define $N=1$ chiral superfields in the Goldstone background,
\begin{equation}
\label{chirality}
\bar{\cal D}_{\dot\alpha} \Phi\ =\ 0\ .
\end{equation}

To understand complex geometry behind this covariant chirality,
and to explicitly solve the constraints \p{chirality_constraint},
\p{chirality}, we consider another, complex parametrization
$\Omega_{\rm L}$ of the coset space $G/H$ (see
\p{real_parametrization}),
\begin{equation}
\label{complex_parametrization}
\Omega_{\rm L}\ =\ \exp i(x_{\rm L}^aP_a +\theta^\alpha Q_\alpha +
\psi^\alpha S_\alpha)
\exp i(\bar\theta_{\dot\alpha}
\bar Q^{\dot\alpha} +\bar\psi_{\dot\alpha}\bar S^{\dot\alpha})\ ,
\end{equation}
where
\begin{equation}
\label{xtrans}
x_{\rm L}^a\ =\ x^a\ -\ i\theta\sigma^a\bar\theta\ -
\ i\psi\sigma^a\bar\psi\ .
\end{equation}
Since the generators $\bar Q^{\dot\alpha}, \bar S^{\dot\alpha}$ and
the Lorentz generators form a (complexified) subalgebra $\tilde H$ of
$G$,  the coordinates $(x_{\rm L}^a, \theta^\alpha, \psi^\alpha)$
of the coset $G/\tilde H$ form a closed subspace under $N=2$
supersymmetry,
\begin{eqnarray}
x'^a_{\rm L}\ &=&\ x_{\rm L}^a\ -
\ 2i\theta\sigma^a\bar\epsilon \ -\ 2i\psi\sigma^a\bar\eta\ , \nn \\
\theta'^\alpha \ &=&\  \theta^\alpha\ +\ \epsilon^\alpha\ ,
\end{eqnarray}
and
\begin{equation}
\psi'^\alpha\ =\ \psi^\alpha\ +\ \eta^\alpha\ ,
\end{equation}
where $\epsilon^\alpha$ and $\eta^\alpha$ are the
first and second supersymmetry transformation parameters.
This implies that we can choose a surface in this space in an
$N=2$ covariant way,
\begin{equation}
\label{surface}
\psi^\alpha\ =\ \psi^\alpha(x_{\rm L}, \theta)\ .
\end{equation}
This equation is equivalent to the holomorphicity condition
\begin{equation}
\label{holomorphicity}
\left({\partial\over \partial \bar\theta_{\dot\alpha} }
\right)_{\rm L}
\psi^\alpha(x_{\rm L}, \theta, \bar\theta)\ =\ 0\ .
\end{equation}
In fact, using \p{holomorphicity}, one can show that the spinor
covariant derivative $\bar{\cal D}^{\dot\alpha}$ becomes a
partial derivative in terms of the complex coordinates
$(x_{\rm L}, \theta, \bar\theta)$,
\begin{equation}
\label{partial}
\bar{\cal D}^{\dot\alpha}\ =\ \left( {\partial\over \partial
\bar\theta_{\dot\alpha} }\right)_{\rm L}\ .
\end{equation}
Thus the holomorphicity condition \p{holomorphicity} is equivalent
to the $N=1$ chirality constraint \p{chirality_constraint}, and
the Goldstone Maxwell superfield $\psi^\alpha$ is a chiral
superfield.  It is now obvious that the general solution to the
covariant chirality constraint \p{chirality} is given by
\begin{equation}
\label{gen_solution}
\Phi\ =\ \Phi(x_{\rm L}^a, \theta^\alpha)\ .
\end{equation}

\subsection{Chiral superspace invariants.}

Having defined the chiral subspace $(x_{\rm L}^a, \theta^\alpha)$,
and chiral superfields covariant under $N=2$ supersymmetry,
we are ready to construct the superspace invariants associated
with chiral superfields.  But first we need a chiral density whose
transformation compensates for the chiral volume transformations
\begin{equation}
\label{measuretrans}
d^4x'_{\rm L}d^2\theta'
\ =\  (1 - 2i\partial^{\rm L}_a\psi\sigma^a\bar\eta)
d^4x_{\rm L}d^2\theta\ .
\end{equation}
One way to obtain such a density is to
take the vielbein superdeterminant $E = {\rm Ber}(E_M{}^A)$ and
then change from the real coordinates $(x, \theta, \bar\theta)$
to the complex coordinates $(x_{\rm L}, \theta, \bar\theta)$,
\begin{equation}
E_{\rm L}\ \equiv\ E\ {\rm Ber}\left({\partial(x_{\rm L},
\theta, \bar\theta)\over \partial(x, \theta, \bar\theta)}\right)\ .
\end{equation}
The density $E_{\rm L}$ transforms correctly,
\begin{equation}
E'_{\rm L}\ =\ (1 + 2i\partial^{L}_a\psi\sigma^a\bar\eta)\  E_{\rm L}\ , \nn
\end{equation}
but it is not chiral:
in the linearized approximation, $E_{\rm L} = 1 + 2i\partial_a\psi
\sigma^a\bar\psi + {\cal O}(\psi^4)$.  The chirality can be restored
with the help of the dimensionless invariants,
\begin{eqnarray}
{\hat E}_{\rm L} &=& E_{\rm L} (1 + {1\over 2}
\bar{\cal D}_{\dot\alpha} \bar\psi_{\dot\beta} \bar{\cal D}^{\dot\alpha}
\bar\psi^{\dot\beta}\ +\ {\cal O}(\psi^4)) \nonumber \\
&=& 1 \ -\ {1\over 4}\bar D^2(\bar W^2) \ +\ {\cal O}(\psi^4)\ .
\end{eqnarray}
The density ${\hat E}_{\rm L}$ transforms correctly, but is
chiral (up to ${\cal O}(\psi^4)$ terms), as can be seen by
its expansion in terms of $W_\alpha$.

Having found the chiral density, we are ready to write
the general $N=1$ superpotential in the Goldstone background.
The coupling is just
\begin{equation}
S_{\rm superpot}\ =\ \int d^4x_{\rm L}d^2\theta \;
{\hat E}_{\rm L}\ P(\Phi)\ ,
\end{equation}
where $P(\Phi)$ is an arbitrary holomorphic function of chiral
superfields.  This coupling is invariant under $N=2$ supersymmetry,
up to ${\cal O}(\psi^4)$.

The discussion of the $N=1$ chiral matter interactions can be
generalized to include gauge multiplets as well.   The
Maxwell gauge superfield is a real $N=1$ superfield, ${\cal A}
(x, \theta, \bar\theta)$, that is a scalar under $N=2$ supersymmetry,
\begin{equation}
\label{gauge_field}
{\cal A}'(x',\theta', \bar\theta')\ =\ {\cal A}(x, \theta, \bar\theta)\ .
\end{equation}
Under gauge symmetry, the superfield ${\cal A}$ transforms as follows,
\begin{equation}
\label{gauge_law}
\delta {\cal A}\ =\ i(\xi(x_{\rm L}^a, \theta) -
\bar\xi(\bar x_{\rm L}^a, \bar\theta))\ .
\end{equation}
where $\xi(x_{\rm L}^a, \theta)$ is a (covariantly) chiral
transformation parameter.

The kinetic term for ${\cal A}$ can be written as an integral over
chiral superspace.  The first step is to construct the chiral
gauge field strength, ${\cal W}_\alpha$, in terms of the Maxwell
superfield, ${\cal A}$, and the Goldstone superfield, $W_\alpha$.
The field ${\cal W}_\alpha$ must be a tensor under gauge symmetry
as well as supersymmetry.  It is
\begin{eqnarray}
{\cal W}_\alpha\ &=&\ i \,\Big[\,
\delta_\alpha^\beta + {1\over 8}\, \delta_\alpha^\beta\,
\bar{\cal D}^2\bar W^2 + \bar{\cal D}_{\dot\beta}
({\cal D}_\alpha W^\beta\bar W^{\dot\beta})
\,\Big]\times \nn\\
&& \quad \Big[\,\bar{\cal D}^2{\cal D}_\beta +
4i({\cal D}_{\beta\dot\alpha}W^\gamma
\bar{\cal D}^{\dot\alpha}\bar W^{\dot\gamma})
\bar{\cal D}_{\dot\gamma} {\cal D}_\gamma
\,\Big] {\cal A}
\ +\ {\cal O}(W^4) \ .
\end{eqnarray}
Then the supersymmetric and gauge-invariant action is just
\begin{equation}
S_{\rm gauge}\ =\ {1\over 4} \int d^4xd^2\theta\ \hat E_L\
{\cal W}^\alpha {\cal W}_\alpha\ +\ h.c.\ ,
\end{equation}
where $\hat E_L$ is the chiral density defined above.

\subsection{Full superspace invariants.}

The kinetic part of the chiral superfield action can also be written
in the Goldstone background.  In flat $N=1$ superspace, the kinetic
action involves a K\"ahler potential, $K(\Phi, \bar\Phi)$,
\begin{equation}
\label{flat_Kahler_potential}
S^{\rm flat}_{\rm kin}\ =\ \int d^4xd^4\theta\ K(\Phi, \bar\Phi)\ ,
\end{equation}
which is defined up to holomorphic K\"ahler transformations
\begin{equation}
\label{flat_Kahler_invariance}
 K'\ =\ K\ +\ \Lambda(\Phi)\ +\ \bar\Lambda(\bar\Phi)\ .
\end{equation}

To generalize \p{flat_Kahler_potential}, \p{flat_Kahler_invariance}
in the Goldstone background, we need a real density $\hat E$ with
the property
\begin{equation}
\label{integration_property}
\int d^4xd^4\theta\ \hat E\ f(x_{\rm L}, \theta)\ =\ 0
\end{equation}
for an arbitrary chiral function $f$.  Expanding the density
$E = 1 + i\partial_a\psi\sigma^a\bar\psi +
i\partial_a\bar\psi\bar\sigma^a\psi +{\cal O}(\psi^4)$
and the function
$f(x_{\rm L}^a, \theta) = (1 - i\psi\sigma^b\bar\psi\partial_b)
f(x^a - i \theta\sigma^a\bar\theta,\theta) + {\cal O}(\psi^4)$,
we see that $E$ does not satisfy \p{integration_property}.  As
above, we can use the dimensionless invariants to define a new
density with the property \p{integration_property},
\begin{equation}
\hat E\ =\  E\, (1 -{\cal D}\psi \bar{\cal D}\bar\psi)
\ +\ {\cal O}(\psi^4)\ .
\end{equation}
The K\"ahler potential part of the matter action is simply
\begin{equation}
\label{Kahler_potential}
S_{\rm kin}\ =\ \int d^4xd^4\theta \ {\hat E}\ K(\Phi, \bar\Phi)
\ ,
\end{equation}
and the K\"ahler potential enjoys the the invariance
\p{flat_Kahler_invariance}.

As discussed above, the matter couplings can be extended to
include $N=1$ gauge multiplets as well.   The kinetic term is
easy to construct in the chiral subspace.  Its associated
Fayet-Iliopoulos term is given by
\begin{equation}
\label{FIterm}
S_{\rm FI}\ =\ \int d^4xd^4\theta\ {\hat E}\ {\cal A}\ ,
\end{equation}
which is invariant under gauge and $N=2$ supersymmetry transformations.

Thus we have seen that all the usual self-couplings of $N=1$ matter
can be extended to the case of partially broken $N=2$ supersymmetry with
the help of the Goldstone-Maxwell multiplet.  A similar result
holds for partially broken $N=2$ supersymmetry with a chiral
Goldstone multiplet \cite{BG}.

\section{Conclusions.}

In this paper we showed that there exists a new Goldstone multiplet
for partially broken $N=2$ supersymmetry, the Goldstone-Maxwell
multiplet.  We found its exact nonlinear supersymmetry transformation
and constructed the invariant Goldstone-Maxwell action.  We also worked
out the first perturbative terms of the $N=1$ matter couplings to the
Goldstone-Maxwell multiplet.  We found that the superspace description
of the Goldstone-Maxwell multiplet requires two constraints,
Eqs.~\p{chirality_constraint} and \p{covariant_b}.  These constraints are
presently on a different footing.   The first, \p{chirality_constraint},
is known in its full form; it has a clear geometrical interpretation
in terms of
$N=1$ chirality preservation.  The second, \p{covariant_b}, is only
known in a perturbative expansion.

The derivation of the second constraint is obscured by two
dimensionless invariants, ${\cal D}_{(\alpha}\psi_{\beta)}$ and
${\cal D}^\alpha \psi_\alpha$.  These invariants can be identified
(at $\theta = 0$) with the gauge field strength, $F_{\alpha\beta}$,
and the auxiliary field, $D$.
It is instructive to compare this situation with that
of the chiral Goldstone multiplet \cite{BG}.  There {\it all}
fields of the Goldstone multiplet were associated with symmetries,
so each had a geometrical interpretation.  For the case at
hand, this suggests that we are missing the Goldstone-type symmetries
associated with the gauge field strength and the auxiliary field
of the Goldstone-Maxwell multiplet.

In fact, the $D$ field of the Goldstone-Maxwell multiplet can be
interpreted as the Goldstone field associated with the following
subgroup of the $SU(2)$ automorphism group of the $N=2$
algebra:
\begin{eqnarray}
\label{U(1)}
\delta\theta^\alpha &=& i\lambda\psi^\alpha \nn\\
\delta\psi^\alpha &=& i\lambda\theta^\alpha\ .
\end{eqnarray}
Under such a transformation, the field $D$ is shifted by the constant
parameter $\lambda$.  This $U(1)$ transformation is a symmetry of the
defining  constraints \p{chirality_constraint}, \p{covariant_b}.
Note that the rest of the automorphism group $SU(2)$ explicitly
breaks these constraints.

If we were to extend $G$ in $G/H$ by \p{U(1)}, we would
eliminate the dimensionless invariant ${\cal D}^\alpha\psi_\alpha$.
However, we would still have to contend with the dimensionless
invariant associated with the gauge field strength, ${\cal D}_{
(\alpha}\psi_{\beta)}$.  This suggests that there exists
an extension of $N=2$ supersymmetry which associates a
Goldstone-like symmetry with this field strength.

\section{Acknowledgements.}

The authors would like to thank E.~Ivanov and J.~Harvey for useful
conversations.  A preliminary attempt to construct a Goldstone theory of
the $N=1$ Maxwell multiplet was discussed by one of us (A.G.) with Viktor
I. Ogievetsky during summer of 1995.  This work was supported by the U.S.
National Science Foundation, grant NSF-PHY-9404057.

\end{document}